\documentclass[12pt]{article}
\usepackage{pic03}
\usepackage{hyperref}
\usepackage{url}
\usepackage{graphicx}

\begin{document}

\title{\bf B, \boldmath $\Lambda_b$ and Charm Results from the Tevatron
\textit{}}
\author{
Farrukh Azfar     \\
{\em Oxford University, Particle Physics Sub-department, 1 Keble Road, Oxford OX1 3RH, UK}}
\maketitle

%
% photograph of author
%  This is where we will insert a photograph. To see what it would look like,
%  uncomment the following lines.
%
\begin{figure}[h]
\begin{center}
% include photograph for proceeding version
\includegraphics
[height=4.5cm]{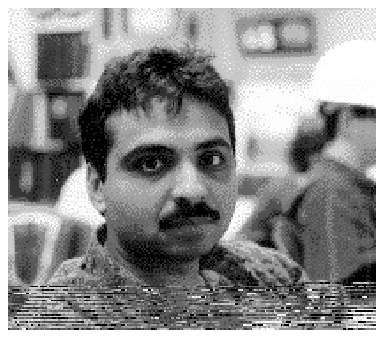}
%
% insert a fixed vertical spacing instead for the ArXiv preprint
%\vspace{4.5cm}
%
\end{center}
\end{figure}

\baselineskip=14.5pt
\begin{abstract}
Recent results on $B_d$, $B_u^{\pm}$, $B_s$, $\Lambda_b$  and Charm hadrons
are reported from $\approx$ 75pb$^{-1}$ and  $\approx$ 40 pb$^{-1}$ of data accumulated at the 
upgraded CDF and D0 experiments at the Fermilab Tevatron $\bar{p}-p$ collider, during Run-II. These include lifetime and mass measurements of $B$ and Charm hadrons, searches for 
rare decays in charm and $B$ hadrons and CP-violation in Charm decays. Results relevant to CP-violation in 
$B$-decays are also reported.
\end{abstract}
\newpage

\baselineskip=17pt

\section{\bf Tevatron $p-\bar{p}$ collider and performance during Run-II}

The Tevatron $p-\bar{p}$ collider is being used for extended data-taking for the second time in 10 years. 
During the period between 1992 and 1996 (Run-I) an integrated luminosity of 
110pb$^{-1}$ was delivered; the goals for the current period starting May 2001 (Run-II) are 2 fb$^{-1}$,  which is a $\times$ 20 increase over Run-I. The upgrades for Run-II consist of a new injection stage delivering more 
protons, an increased $\bar{p}$ transfer efficiency and a $\bar{p}$ recycler (undergoing commisioning) that uses 
remaining $\bar{p}$ from the  previous store. A table of Run-I and Run-II operating parameters is given in table 
~\ref{tevtab}. The peak luminosity, though improving, is still  $\times$ 4 below target.

\begin{table} [h]
\centering
\caption{ \it Tevatron Performance Improvement Run-I vs. Run-II.
}
%\vskip 0.75 in
\begin{tabular}{|l|c|c|c|c|} \hline & Collision Rate & Bunches & Center of Mass Energy & Peak Luminosity \\
\hline
\hline
Run-I:  &  3.5$\mu$s (Run-I) & 6x6 & 1.8 TeV/c$^2$ & 2.4x10$^{31}$ \\
Run-II: &  396 ns & 36x36 & 1.96 Tev/c$^2$ & 4.4x10$^{31}$ \\
\hline
\end{tabular}
\label{tevtab}
\end{table}

\section{\bf $B$ Physics at Hadron Colliders: The CDF and D0 Detectors}
The $b\bar{b}$ production cross section $\sigma(b\bar{b})$ is $\approx$ $150\mu$b at $p-\bar{p}$ at the 
Tevatron, 1 nb at the $\Upsilon(4s)$ and 7 nb at the $Z_0$. All $B$-hadrons are produced at the 
Tevatron (unlike the $B$ factories), the drawback being the inelastic cross section which is 
1000$ \times \sigma(b\bar{b})$ making online data selection crucial. 

The CDF~\cite{cdfdetector} and D0 detectors~\cite{d0detector} have been described elsewhere. In order to utilize the high $b\bar{b}$ production cross section clean signatures of $B$ hadron decays must be used 
when selecting data online. During Run-I CDF used the clean signatures of $B \to J/\psi \to \mu^+ \mu^-$ and the decays of $B$ hadrons to high transverse momentum ($P_T$ ) leptons. Now the long lifetime
of $B$ hadrons is being utilized at CDF, events containing $\ge$2 tracks with high impact parameters ($d_0$) 
consistent with being daughters of $B$ hadrons are selected using the Silicon Vertex Trigger, events selected in 
this way are categorized as ``hadronic $B$ trigger'' or ``displaced track trigger'' events. 
Another trigger used for $B$ selection selects events requires the presence of a single high $P_T$ lepton and 
a high $d_0$ track, this is called the ``lepton+SVT trigger '' or ``lepton+displaced track trigger''.
The high $d_0$ triggers have allowed CDF to significantly advance its $B$ and charm physics capability, 
now CDF can reconstruct $B$ decays to fully hadronic final states as well as use the conventional $J/\psi$ and
high $P_T$ lepton triggers.

The D0 experiment utilizes the $\mu^+ \mu^-$ signature to select $J/\psi$s (which may be prompt or long-lived---coming from $B$ hadrons). D0 triggers also use $B$ decays to high $P_T$ leptons. With the advent of
a magnetic field and a completely new tracking system D0 has acquired a whole new capability in $B$ physics
and several $B \to J/\psi X$ decay modes have been observed during Run-II. The D0 experiment is on 
its way to introducing a hadronic $B$ trigger that uses online silicon pattern recognition to select tracks with 
high $d_0$, this will be a welcome addition to D0's already enhanced capability in $B$ physics.

\section{\bf Physics Results: Testing Heavy Quark Expansion: Lifetimes of $B$ Hadrons at CDF and D0}
Precise measurements of lifetimes of $B$ hadrons allow tests of the Heavy Quark Expansion (HQE) 
which predicts the following hierarchy for $B$ hadron lifetimes: $\tau_{B_c^{\pm}} < \tau_{\Lambda_b} < \tau_{B_d} \approx \tau_{B_s} < \tau_{B_u^{\pm}}$. Both CDF and D0 are working toward precision tests of this theory.

\subsection{\bf Charged to Neutral Lifetime Ratio of $B$ Hadrons, and $B_s$ Lifetime from fully 
reconstructed decays}

The ratio of the lifetimes of $B_u^{\pm}$ to $B_d$ has been measured at CDF, and the $B_u^{\pm}$ lifetime
has been measured at D0 using fully reconstructed $B_d \to J/\psi K^*$ and $B_u^{\pm} \to J/\psi K^{\pm}$
decays. The decays are fully reconstructed, the invariant mass and proper decay length ($c\tau$) 
distributions of the $B$s are calculated from an un-binned log-likelihood function determining both mass and lifetime in a single fit. The result is $1.11 \pm 0.09$ at CDF using 70 pb$^{-1}$ of data, whereas the $B_u^{\pm}$ 
lifetime has been measured to be $1.76 \pm 0.24$ ps at D0. Both these measurements test HQE and are consistent with the more accurate measurements at BaBar and Belle.

\subsection{\bf $B_s$ Lifetime from the fully reconstructed decay $B_s \to J/\psi \phi$, $J/\psi  \to \mu^+ \mu^-$ and $\phi \to K^+ K^-$ }

\begin{figure}[htbp]
  \centerline{\hbox{ \hspace{0.2cm}
    \includegraphics[width=6.5cm,height=5.0cm]{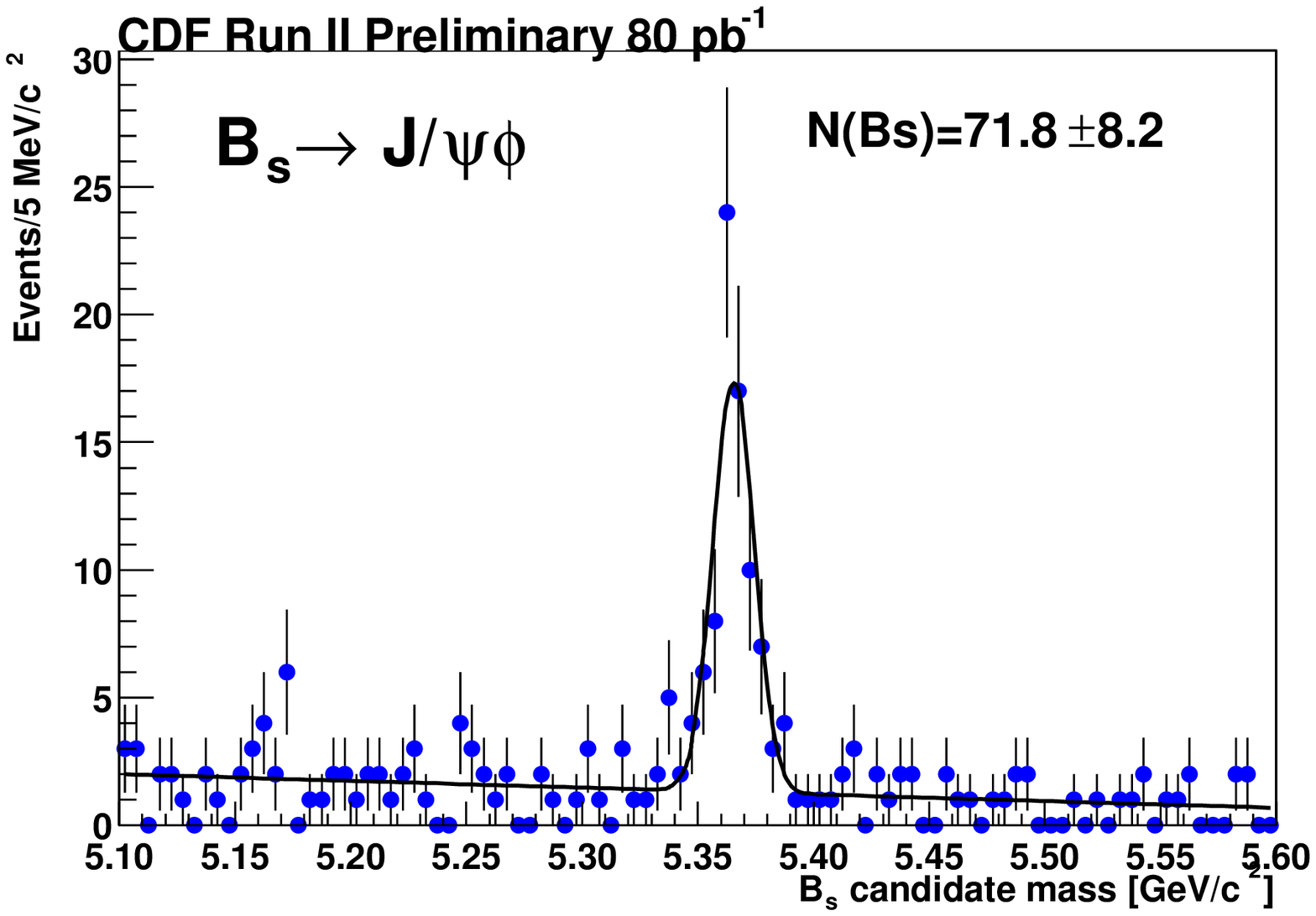}
    \hspace{0.3cm}
    \includegraphics[width=6.5cm,height=5.0cm]{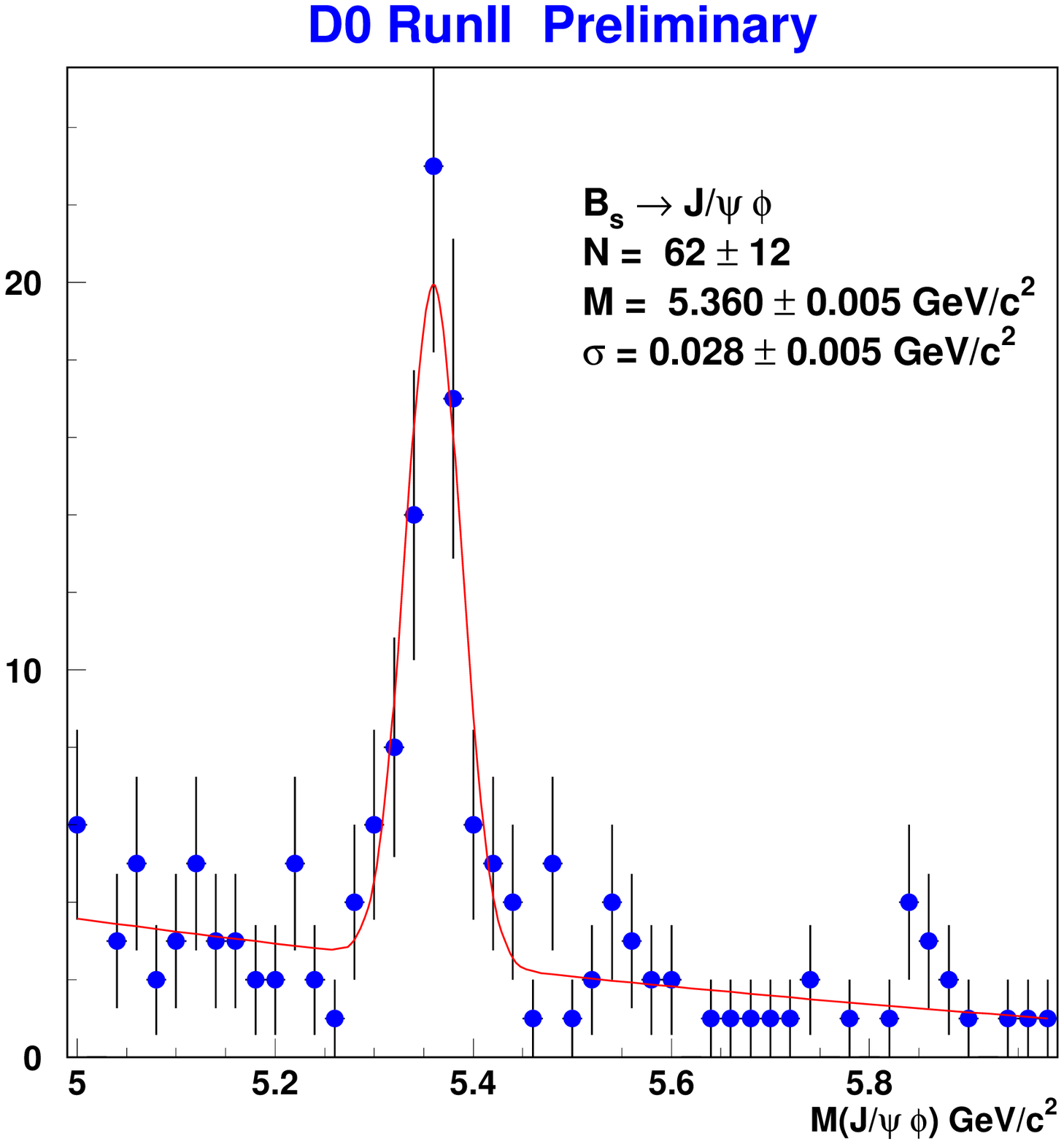}
    }
  }
 \caption{\it
      $B_s \to J/\psi \phi$ with $J/\psi \to \mu^+ \mu^-$ and $\phi \to K^+ K^-$ at CDF (72 events) and D0 (61 events)
    \label{bsjpsiphi} }
\end{figure}

The world's largest number of  fully reconstructed $B_s$ decays has been at the Tevatron since Run-I. CDF and 
D0 has successfully reconstructed this decay using data from their  $J/\psi \to \mu^+ \mu^-$ trigger during Run-II, the 
reconstructed signals are shown in Fig.~\ref{bsjpsiphi}. A measurement of the lifetime is underway at D0, CDF 
has measured a ratio of $\frac{\tau{B_s}}{\tau{B_d}}= 0.89 \pm 0.15$ based on 70pb$^{-1}$ collected during Run-II.

Strictly speaking this ratio isn't what should be tested, since the CP composition of the final
state is not known. The final state is a mixture of CP eigenstates, and the $B_s$ CP even and odd eigenstates can 
have a difference in lifetime of upto 10 \% as predicted by theory, by fitting a single lifetime in this mode an 
``average'' lifetime has been determined. The relationship between splitting in width (lifetime) of the $B_s$ CP 
eigenstates and the $B_s$ mixing parameter is known in the Standard Model, therefore a measurement of the 
width (lifetime) difference and mixing parameter provides a test of new Physics. It is planned to measure the 
lifetime difference of the $B_s$ CP eigenstates by utilizing angular variables to disentangle the two CP eigenstates 
and fitting two lifetimes, this approach will become viable with higher statistics. An accuracy of 5 \% in determining 
the lifetime splitting is expected at CDF using 4000 $B_s \to J/\psi \phi$ decays. 

A clean measurement of the width difference $(\frac{\Delta \Gamma}{ \Gamma})_{Bs}$, can be made by 
measuring a single lifetime in a decay of the $B_s$ to a CP eigenstate {\it e.g.} $B_s \to D_s^+ D_s^-$ and 
$B_s \to K^+ K^-$. This approach is planned as well.

\subsection{\bf  Measurement of the $\Lambda_b$ Lifetime }
\begin{figure}[htbp]
  \centerline{\hbox{ \hspace{0.2cm}
    \includegraphics[width=6.5cm,height=5.0cm]{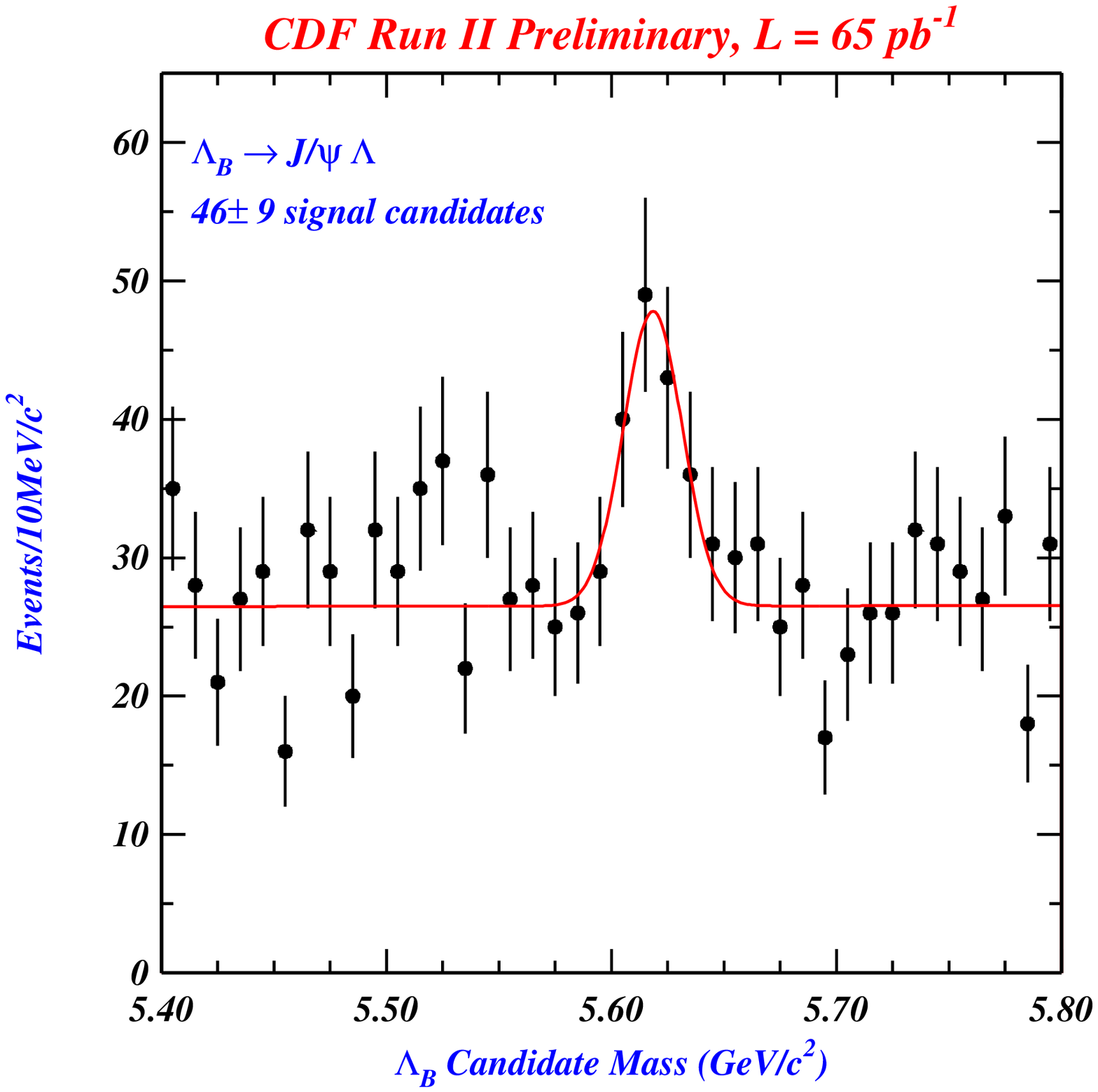}
    \hspace{0.3cm}
    \includegraphics[width=6.5cm,height=5.0cm]{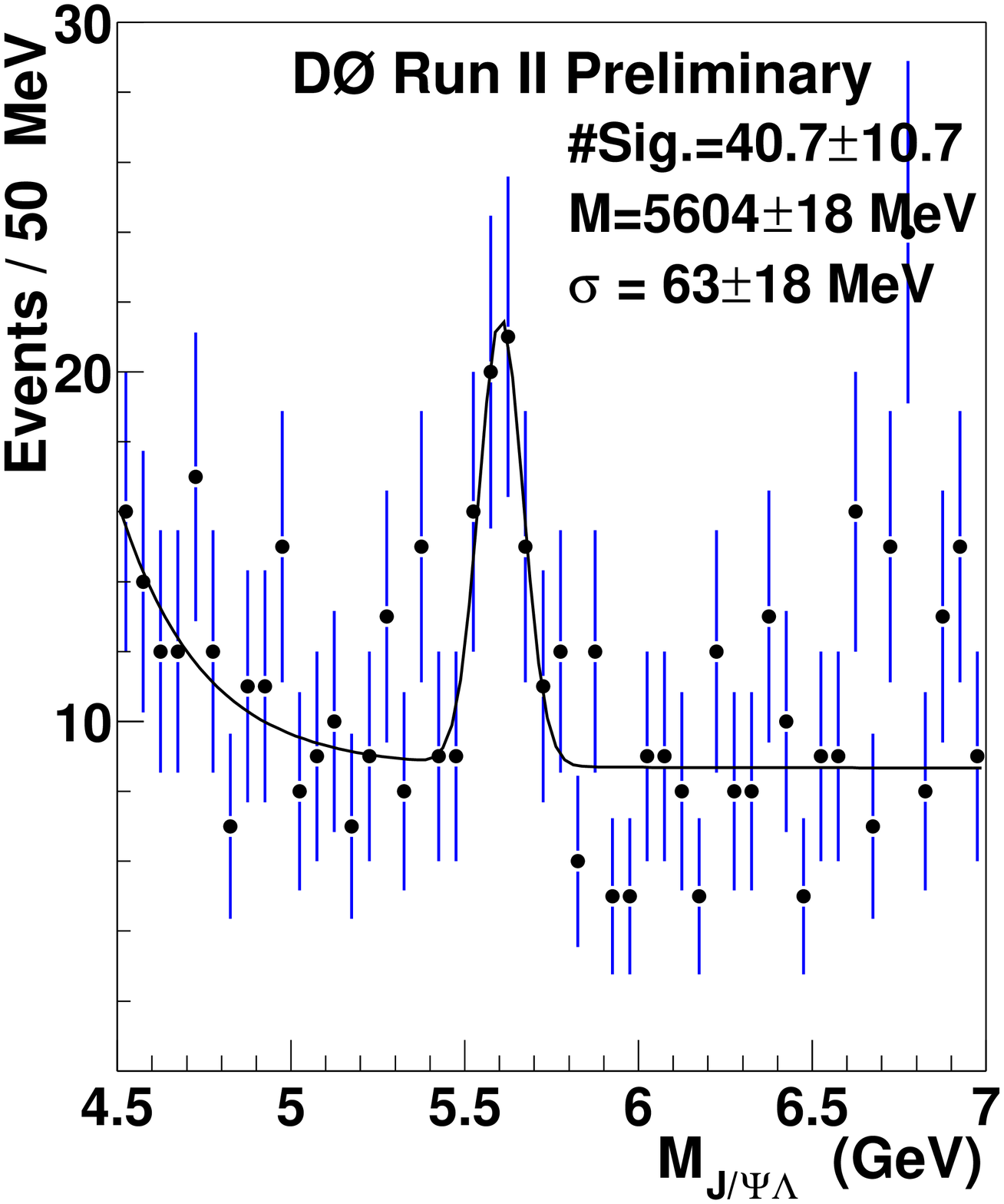}
    }
  }
 \caption{\it
      CDF (left) and D0 (right) reconstruction of $\Lambda_b \to J/\psi \Lambda$.
    \label{lambda_b} }
\end{figure}

The $\Lambda_b$ has been reconstructed at both CDF and D0 in various modes, Fig.~\ref{lambda_b} 
shows the reconstruction in the fully hadronic mode at CDF and in $\Lambda_b \to J/\psi \Lambda$ at D0.

\begin{figure}[htb]
\includegraphics[width=13cm, height=6.0cm]{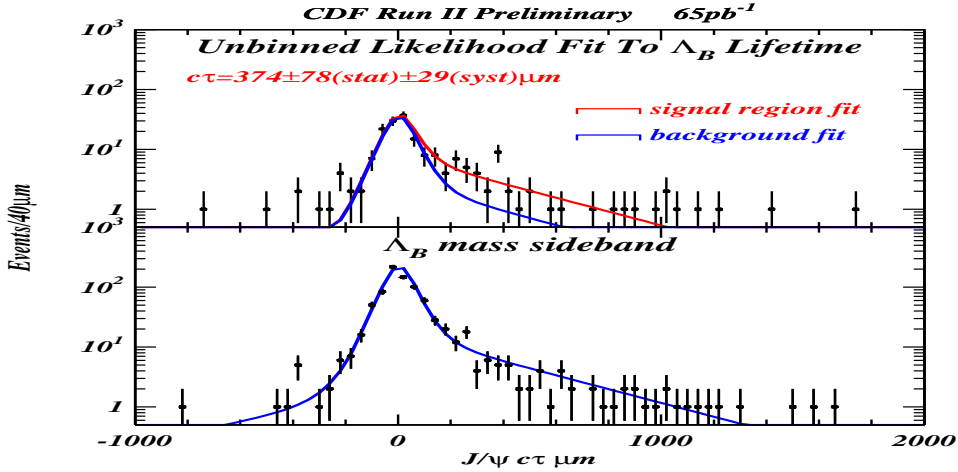}
 \caption{\it
Lifetime distribution of $\Lambda_b \to J/\psi \Lambda$ decays at CDF in micro-meters.
    \label{lambda_b_life} }
\end{figure}

CDF has also reconstructed $\Lambda_b$ in the decays $\Lambda_b \to J/\psi \Lambda$ (53 events) and 
$\Lambda_b \to \Lambda \ell \nu_{\ell}$ (640 events) and in the purely hadronic decay mode 
$\Lambda_b \to \Lambda_c^{\mp} \pi^{\pm}$ with $\Lambda_c \to pK\pi$. A lifetime measurement has just 
been completed at CDF using the fully reconstructed decay $\Lambda_b \to J/\psi \Lambda$ with a result 
$\tau_{\Lambda_b}=1.25 \pm 0.26 (stat) \pm 0.1(syst)$ ps shown in Fig.~\ref{lambda_b_life}. Work on D0's $\Lambda_b$ lifetime currently underway. 

The decays $\Lambda_b \to \Lambda_c^{\pm} \ell \nu_{\ell}$ and $\Lambda_b \to \Lambda_c^{\pm} \pi^{\mp}$ are selected using a trigger with $d_0$ cut, the resulting biases in the $c\tau$ distribution have to be understood before a lifetime measurement from these modes can be completed and CDF expects this will be done soon.

\begin{figure}[htb]
\includegraphics[width=13cm, height=6.0cm] {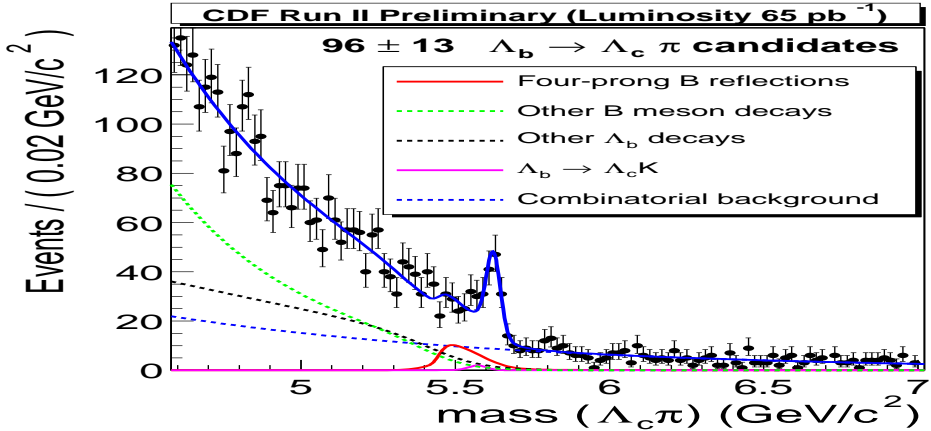}

 \caption{\it
 Invariant Mass distribution of the purely hadronic $\Lambda_b$ decay mode $\Lambda_b \to \Lambda_c^{\pm} \pi^{\mp}$ at CDF.
    \label{lambda_pure_hadronic} }
\end{figure}

\section{\bf Charm Physics: $D_s^{\pm}-D^{\pm}$ mass difference}
The first CDF Run-II publication ~\cite{dsmassdiff} was a measurement of the mass difference 
$\Delta M=M_{D_s^{\pm}}-M_{D^{\pm}}$. Both the $D_s^{\pm}$ and $D^{\pm}$ decay to 
$\phi \pi^{\pm}$ with $\phi \to K^+ K^-$ with almost identical kinematics. Using data selected by the displaced-track hadronic trigger 2400 $D_s^{\pm}$ and 1600 $D^{\pm}$ were reconstructed using only 11.6 pb$^{-1}$ of data. 
The measurement of $\Delta M = 99.28 \pm 0.43 (stat) \pm 0.27 (syst) $ MeV/c$^2$ is consistent with the current world average ~\cite{PDGallgemein} of 99.2 $\pm$ 0.5 MeV/c$^2$. 

\section{\bf  Rare Decays: The Search for the Flavour Changing Neutral Current Decay $D \to 
\mu^+ \mu^-$ and $B_s^0 \to \mu^+ \mu^-$}

The standard model predicts a branching ratio of $O(10^{-13})$ for the decay 
$D^0 \to \mu^+ \mu^-$ via second order weak interactions. Some R-parity violating SUSY 
models predict branching ratios $\le O(10^{-6})$ ~\cite{smfcnc}. CDF has searched for 
$D^0 \to \mu^+ \mu^-$ decays using hadronic trigger data and $D^0 \to \pi^+ \pi^-$ decays which have 
almost identical acceptance and kinematics to $D^0 \to \mu^+ \mu^-$ . The probability of a $\pi^{\pm}$ faking a 
$\mu^{\pm}$  must be calculated, unambiguously identified pions are obtained using the 
decay chain ${D^*}^{\pm} \to D^0 \pi^{\pm}$, $D^0 \to K^{\mp} \pi^{\pm}$, the charge of the $\pi^{\pm}$ 
from the $D^{* \pm}$  determines the flavour of the $D^0$  and distinguishes the $K^{\pm}$ from the $\pi^{\pm}$. 
The number of times a $\pi^{\pm}$ is reconstructed as a $\mu^{\pm}$ is determined after which 
$D^0 \to \mu^+ \mu^-$ are reconstructed and expected number of $D^0 \to \pi^+ \pi^-$ decays faking  $D^0 \to \mu^+ \mu^-$ is subracted, 0 events are found in a 2$\sigma$ search window. A limit for this branching ratio $\le 2.4 \times 10^{-6}$ is calculated at 90 \% confidence level, better than the best published limit of $4.1 \times 10^{-6}$~\cite{beatrice}~\cite{PDGallgemein}.

CDF has done a similar analysis of the decay $B_s \to \mu^+ \mu^-$ using 113 pb$^{-1}$ of Run-II data. 
Standard Model prediction for the branching ratio is $3.8 \pm 1 \times 10^{-9}$. Various SUSY 
models~\cite{bsmumuXSM1} allow for an enhancement by a factor of upto $\times 10^3$, areas of m-SUGRA 
space that overlap those predicting deviations of the $g_{\mu}$ from 2 are roughly consistent~\cite{bsmumuXSM2}
with recent experimental measurements~\cite{mugyromag}. CDF's measurement yields limits 
$BR(B_s \to \mu^+ \mu^-)$ $< 9.5 \times 10^{-7}$ and $1.2 \times 10^{-6}$ at the 90 \% and 95 \% 
confidence intervals respectively---a factor of 2 better than the best previous measurement~\cite{PDGallgemein}.

\section{\bf  CP Violation in Charm Decays}
The Standard model prediction for CP violation in charm decays is of order 0.1-1 \%.
Since $c$ and $u$ quarks do not couple to $t$ quarks, box diagram contributions to
mixing in charm are tiny, and so CP violation in Charm decays is almost entirely due to interference in decay 
(direct CP violation). A search for CP violation in charm decays has been done at CDF. Rates of decays of $D^0$ 
and $\bar{D^0}$ decaying to the CP eigenstates $f=K^+K^-$ and $f=\pi^+ \pi^-$ are measured. The flavour of 
the $D^0$ is tagged as described in section 5, and $D^0 \to \pi^+ \pi^-$ and $\to K^+ K^-$ decays 
are reconstructed and counted and the asymmetry
\begin{equation}
A_{CP}=\frac{\Gamma(D^0 \to f)-\Gamma(\bar{D^0} \to f)}{\Gamma(D^0 \to f)+\Gamma(\bar{D^0} \to f)}
\label{charmcpass}
\end{equation}
for each mode is calculated. The results are $A_{CP}(D^0 \to \pi^+ \pi^-) = 2.0 \pm 1.7 (stat) \pm 0.6 (syst)$ \% 
and $A_{CP}(D^0 \to K^+ K^-) = 3.0 \pm 1.9 (stat) \pm 0.6 (syst)$ \%, consistent with both the world averages 
of $0.5 \pm 1.6$ \% and $2.1 \pm 2.6$ \%  ~\cite{PDGallgemein} and better than the most recent (2001) CLEO 
results of  $0.0 \pm2.2 \pm 0.8$ \% and $1.9 \pm 3.2(stat) \pm 0.8 (syst)$\% respectively~\cite{cleocharmcp}. 

As a check of possible biases in counting, the ratios of branching ratios: 
$\frac{\Gamma(D^0 \to K^+ K^-)}{\Gamma(D^0 \to K^{\pm} \pi^{\mp})}$ and 
$\frac{\Gamma(D^0 \to \pi^+ \pi^-)}{\Gamma(D^0 \to K^{\pm} \pi^{\mp})}$ were also calculated 
and found to be $9.38 \pm0.18 (stat) \pm 0.10(syst) $\% and $3.686 \pm 0.076 (stat) \pm 0.036 (syst)$\%  
respectively. These compare well with the measurements at FOCUS~\cite{focusratofbr} $9.93 \pm 0.14 (stat) \pm 0.14(syst)$\% and $3.53 \pm0.12 (stat) \pm 0.06(syst)$\%.

\section{\bf Towards CP violation in $B$-hadron decays and $B_s$ mixing}
In Run-I the CDF was able to competetively measure the $B_d$ mixing parameter 
$(\frac{\Delta M}{\Gamma})_{B_d}=x_d$ and also perform a 2$\sigma$ measurement of the CP asymmetry 
in the decay $B_d \to J/\psi K_S$ ($\sin 2\beta$)~\cite{cdfsin2beta}. The Run-I measurement was 
$\sin 2\beta = 0.79 \pm 0.39 (stat) \pm 0.16(syst)$, BaBar and Belle already have measurements of $0.76 \pm 0.067(stat) \pm 0.034(syst)$ and $0.733 \pm 0.057 (stat) \pm 0.028(syst)$ respectively~\cite{bellesin2beta} 
~\cite{babarsin2beta}. 
With $\times$40-50 more decays expected when 2 fb$^{-1}$ have been accumulated, CDF's precision should be 
$\delta (\sin 2 \beta) \approx 0.05$, D0 should have similar statistics. Clearly D0 and CDF cannot compete with  the 
$B$-factories $\sin 2\beta$ measurement, but $\sin 2\beta$ will be measured as a benchmark, and a test of various flavour tagging schemes.

Various tagging schemes are under examination at CDF; including jet-charge, opposite and same-side tagging and using time of flight to identify $K$s. A final number for the statistical power {\it i.e.} $\epsilon D^2$ has not yet been calculated using data. 

\subsection{\bf Measurement of $\sin 2\gamma$ using $B_d \to \pi^+ \pi^-$ and $B_s \to K^+ K^-$ }
\begin{figure}[htb]
\includegraphics[width=13cm, height=6.0cm] {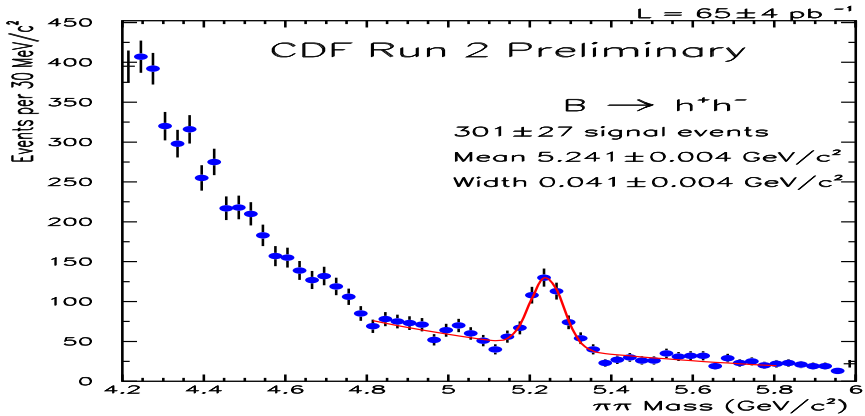}

 \caption{\it
 Invariant Mass distribution of all $B \to h^+ h^-$, decays at CDF. Both tracks are assigned the mass of a $\pi$.
    \label{btohh} }
\end{figure}

Both tree and penguin graphs contribute to $B_d \to \pi^+ \pi^-$  and $B_s \to K^+ K^-$ with the tree
dominating in the former and the penguin in the latter. Without the penguin contributions the CP asymmetry 
($A_{CP}$) in  $B_d \to \pi^+ \pi^-$ is proportional to the CKM quantity $\sin 2(\gamma + \beta)$  and 
$A_{CP}$ in $B_s \to K^+ K^-$  is proportional to $\sin 2\gamma$. Assuming {\bf SU(3)} symmetry 
and interchanging  $s$ and $d$, the hadronic matrix element penguin to tree ratios are the same,  the mixing and 
decay induced $A_{CP}(t)$ are functions of $\sin 2\gamma$, $\sin 2\alpha$, $\sin 2\beta$, the ratio of the hadronic matrix element amplitudes and the phase of this ratio. A measurement of the $A_{CP}$ thus determines 
$\sin2 \gamma$ and  $\sin 2\alpha$.
Before measuring this asymmetry the various $B_d \to h^+ h^-$ and $B_s \to h^+ h^-$ decays must be separated. 
Reconstructing $B_d \to \pi^+ \pi^-$ without clear hadron identification leads to a very 
broad peak in which the individual modes  $B_d \to K^{\pm} \pi^{\mp}$, $B_s \to K^{\pm} \pi^{\mp}$, $B_d \to \pi^{+} \pi^{-}$ and $B_s \to K^{+} K^{-}$ 
are indistinguishable. These can be separated at CDF utilizing $\frac{dE}{dx}$ using drift 
chamber charge deposition and kinematical variable separation. CDF has reconstructed  
$39 \pm 14$ $B_d \to \pi^+ \pi^-$ and $90 \pm 17$ $B_s \to K^+ K^-$ decays, the latter is a first observation. 
The invariant mass distribution of  all $B$ hadrons decaying to $h^+h^-$ is shown in figure ~\ref{btohh}. 

As a check the ratio of branching ratios $\frac{\Gamma(B_d \to \pi^+ \pi^-)}{\Gamma(B_d \to K^{\pm} \pi^{\mp})}$ has been measured, the result $0.26 \pm 0.11 (stat) \pm 0.055 (syst) $ is consistent with the world-average 
$0.253 \pm 0.064$~\cite{PDGallgemein}. The CDF experiment expects to be able to measure $\gamma$ to an accuracy of $\sigma(\gamma) \approx 10$ degrees.

\subsection{\bf Measurement of the $B_s$ Mixing parameter $x_s=\frac{\Delta M_{B_s}}{\Gamma_{B_s}}$}

\begin{figure}[htb]
\includegraphics[width=13cm, height=6.0cm]{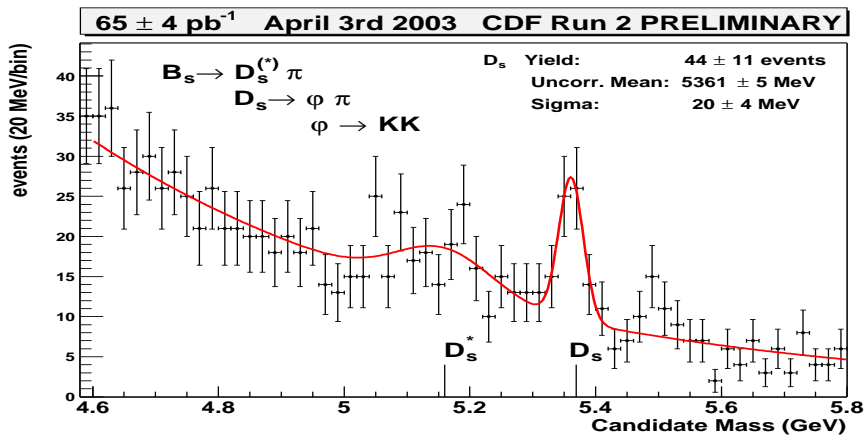}
 \caption{\it
First observation of $B_s \to D_s^{\pm} \pi^{\mp}$, made at CDF.
    \label{bsdspi} }
\end{figure}

The measurement of the $B_s$ mixing parameter $x_s = \frac{\Delta M_{B_s}}{\Gamma_{B_s}}$ is one of 
the major goals of the Tevatron during Run-II.  An observation of the flagship mode for measuring 
$x_s$, $B_s \to D_s^{\pm}\pi^{\mp}$ has been made at CDF. In addition to this mode $B_s$ mixing can also be measured using modes such as $B_s \to \mu^{\pm} \nu_{\mu}D_s^{\mp}$ and $B_s \to e^{\pm} \nu_{e}D_s^{\mp}$, however the vertex resolution in these decays is worse due to the missed neutrino. If a single 
$B_s$ lifetime is fit in any of these or any flavour specific mode the relation between the fit lifetime $\tau_{fit}$ and 
the CP odd and even lifetimes $\tau_{CP+}$, $\tau_{CP-}$  is $\tau_{fit} = \frac{(\tau_{CP+}^2 + \tau_{CP-}^2)}{(\tau_{CP+}+ \tau_{CP-})}$, which can be used for a measurement of $\Delta \Gamma_{B_s}$ and to provide a useful 
constraint for the two-lifetime fit with $B_s \to J/\psi \phi$ described earlier. The first observation of 40 $B_s \to D_s^{\pm}\pi^{\mp}$  decays has been made using hadronic trigger data at CDF, shown in Fig.~\ref{bsdspi}. Also 309 
$B_s \to \mu^{\pm} \nu_{\mu}D_s^{\mp}$ and 245 $B_s \to e^{\pm} \nu_{e}D_s^{\mp}$ have been observed using lepton+displaced track trigger data.

\section{\bf Conclusions}
Both CDF and D0 are in the first phase of data taking ($\int Ldt <200$pb$^{-1}$) which will test HQE 
with $\Lambda_B$ and $B_s$ lifetimes, and yield limits for CP violation and rare decays in $B$ 
and charm decays. Tagging and lifetime measurement techniques (for hadronic trigger data) will also be tested. 
In the next phase ($200< \int L dt <500$pb$^{-1}$) limits on $B_s$ mixing will be set and CP violation searches 
in the $B$ system will be done. In the final phase ($500< \int Ldt <2000$pb$^{-1}$)  $\Delta \Gamma_{B_s}$,
$x_s$, and the CKM angle $\gamma$ will be measured and finally a search for unexpectedly large CP violation in $B_s \to J/\psi \phi$ will be pursued.

\end{document}